\documentclass[runningheads]{llncs}
\usepackage{graphicx}
\begin{document}
\title{Endotracheal Tube Detection and Segmentation in Chest Radiographs using Synthetic Data}

\author{Maayan~Frid-Adar\inst{1}\orcidID{0000-0002-8246-1411} \and
Rula~Amer\inst{1}\orcidID{0000-0002-5630-827X} \and
Hayit~Greenspan\inst{1,2}\orcidID{0000-0001-6908-7552}}

% \author{******** \\ *****}

\institute{RADLogics Ltd., Tel-Aviv, Israel\and
Department of Biomedical Engineering, Tel Aviv University, Tel Aviv, Israel
\email{\{maayan,rula,hayit\}@radlogics.com}\\
\email{hayit@eng.tau.ac.il}}

% \institute{*** \\ *** \\ *** \\ ***}

%
\maketitle              % typeset the header of the contribution
\begin{abstract}
Chest radiographs are frequently used to verify the correct intubation of patients in the emergency room. Fast and accurate identification and localization of the endotracheal (ET) tube is critical for the patient. In this study we propose a novel automated deep learning scheme for accurate detection and segmentation of the ET tubes. Development of automatic systems using deep learning networks for classification and segmentation require large annotated data which is not always available. Here we present an approach for synthesizing ET tubes in real X-ray images.  We suggest a method for training the network, first with synthetic data and then with real X-ray images in a fine-tuning phase, which allows the network to train on thousands of cases without annotating any data.
The proposed method was tested on 477 real chest radiographs from a public dataset and reached AUC of 0.99 in classifying the presence vs. absence of the ET tube,  along with outputting high quality ET tube segmentation maps.

\end{abstract}
\section{Introduction}
The American College of Radiology recommends acquisition of chest radiographs following intubation, to ensure proper positioning of inserted tubes, for patients in the Intensive Care Unit (ICU)  \cite{godoy2012chest}. This is justified by studies, such as \cite{trotman2003radiology}, which show that following intubation, physical examination identified tube malposition in only 2\% to 5\% of patients, whereas the radiograph revealed suboptimal positioning in 10\% to 25\%.
The ideal endotracheal (ET) tube position is in the mid trachea if the patient’s head is in the neutral position. Malposition of the ET tube can cause serious complications if not detected, especially where the tube is too low and selective bronchial intubation occurs. Such complications include a segmental or complete collapse of the contralateral lung, pneumothorax and atelectasis \cite{trotman2003radiology}.

Using the acquired radiographs, Computer-Aided Detection (CAD) systems can assist physicians in automatic detection of  the ET tubes. Previous studies used classical approaches to determine seed points followed by a line tracking algorithms \cite{ramakrishna2012improved,chen2016endotracheal}.
A more recent study used a convolutional neural network (CNN) classification system for the presence or absence identification of the ET tube,  with reported area under curve (AUC) of 0.99; and a second classification network for identification of low vs normal positioning of the ET tube, with AUC of 0.81 \cite{lakhani2017deep}. The above studies used private datasets of portable chest X-ray images with a relatively small amount of cases: 64 \cite{ramakrishna2012improved} and 87 \cite{chen2016endotracheal} were used for the classical approaches; 300 cases were used for the CNN based solution \cite{lakhani2017deep}.
%- 64 and 87 cases, respectively, for developing and testing of the classical approaches \cite{ramakrishna2012improved,chen2016endotracheal}. And 300 cases overall for each of the classification problems in the networks based study (with 60 cases each for testing). 

Collecting and labeling chest radiographs for presence of ET tubes requires  collaboration with hospitals and data extraction methods. For the ET tube detection and  ET tube segmentation annotation,   expert physicians are needed.
In this paper, we present an innovative solution for the task of detection and segmentation of ET tubes in chest radiographs, in the {\it{scenario of limited expert labeled data}}: We use a {\it{public dataset of chest radiographs}} \cite{wang2017chestx} which allows us to collect a large data of normal and ET tube examples required for training a deep learning network. We then {\it{synthesize ET tubes}} on top of the X-ray images to generate ground truth data for the ET tube segmentation. Finally, we present a combined CNN for ET tube detection and segmentation in chest radiographs showing promising results.

\begin{figure}[!t]
\centering
\includegraphics[width=4.7in]{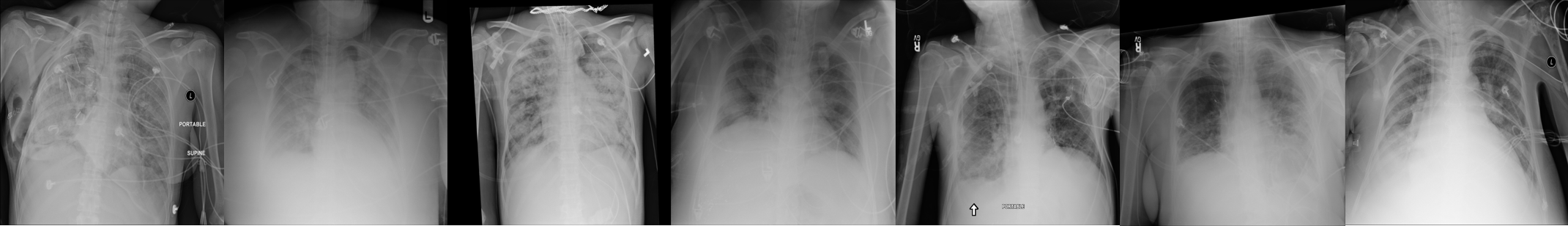}
\caption{Examples from the NIH public dataset \cite{wang2017chestx}}
\label{fig:nih_examples}
\end{figure}

\begin{figure}[!t]
\centering
\includegraphics[width=2.8in]{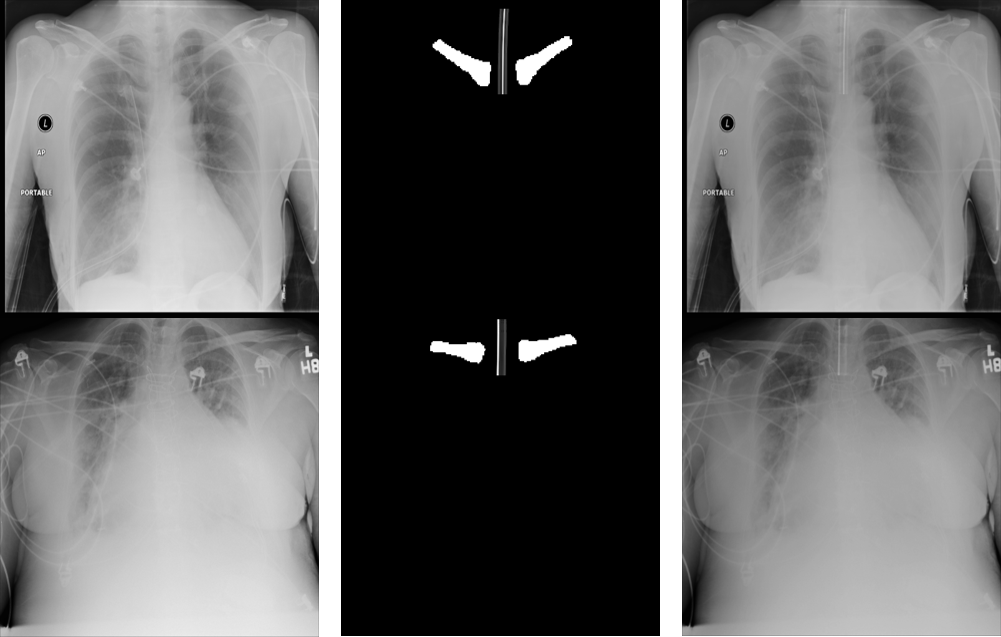}
\caption{Original X-ray images (left column), Clavicles segmentation and synthesized ET tube (middle column) and Synthesized ET tubes blended into the original X-ray (right column) }
\label{fig:ETT_synth}
\end{figure}

\section{Methodology}
\label{sec:methodology}

% Training deep neural networks requires large amount of training images. 
%Collecting medical images with annotations is a long process of collaborative
%effort between researches and radiologists. 
%For that reason, 
%A growing number of public medical datasets have become available in recent years. %However, annotated medical data is still lacking for many applications. In such cases, %creative solutions such as using synthetic data, are often needed.
In this study, we apply a technique to insert synthetic ET tubes as an overlay to the original X-ray images
taken from a publicly available dataset of chest radiographs \cite{wang2017chestx} (hereon will be called the NIH dataset). This dataset contains over 100,000 frontal view images,
many of them coming from ICU patients. While annotations are provided
for 14 lung diseases, no annotations exist for the presence of ET tubes
(or other tubes). A few sample images from the NIH dataset are shown in Figure \ref{fig:nih_examples} -  the cases have high variability and many have poor image quality. We only used cases in Anterior-Posterior (AP) positioning to simulate intubated patients.

In the first step of our proposed solution, we propose a technique to generate  new images with ground truth ET tube segmentation masks.
The new image set we form will be used in a follow-up step, for training a combined CNN for detection and segmentation of ET tubes in chest
radiographs.

\subsection{Generating Synthetic Data}
\label{ssec:generatesynthdata}
Generating the synthetic ET tubes over real X-ray images includes the following main steps as shown in Fig. \ref{fig:ETT_synth}: a) Selection of cases from the NIH dataset that do not contain ET tubes but may include other tubes (such as nasogastric (NG) tube, drainage tubes, catheters); b) Segmentation of the clavicles in order to localize the synthetic ET tube in the trachea area; c) Blending of generated synthetic ET tubes onto real X-ray images.

% \textbf{Clavicles Segmentation:}
\subsubsection{Clavicles Segmentation:}
%\label{sssec:claviclesseg}
ET tubes are inserted into the trachea to allow artificial ventilation of the lungs. X-ray images are mostly aligned with the trachea located between the clavicles. Therefore, correct segmentation of the clavicles assists in placing the synthetic ET tube in the trachea area. In \cite{frid2018improving} a methodology for organ segmentation within Chest radiographs was presented, and shown to outperform alternate schemes, when tested on a common benchmark of 247 chest radiographs from the JSRT dataset, with ground-truth segmentation masks from the SCR dataset \cite{van2006segmentation}. The architecture proposed is based on a modified U-Net based architecture, in which  pre-trained encoder weights were used, based on VGG16. In the current work, we use a similar scheme: For training we input $224\times 224$  images, each normalized by its mean and standard deviation. We train the single-class segmentation model using Dice loss and threshold the output score maps to generate binary segmentation masks of the clavicles structure. This model gives us Dice coefficient score of 93.1\% and Mean average contour distance of 0.871 mm.

\begin{figure}[!t]
\centering
\includegraphics[width=4.5in]{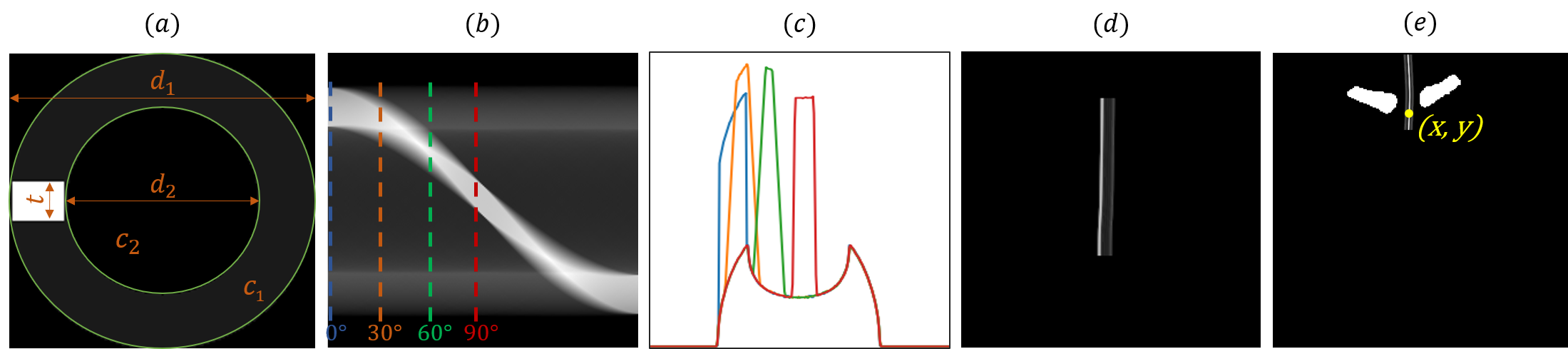}
\caption{Realistic ET tubes generation: (a) ET tube cross profile; (b) 2D projection of the tube; (c) Sampled profile at different angles; (d) ET tube profile drawing; (e) Drawing the ET tube relative to the location of the clavicles (using middle point x and lower point y)}
\label{fig:generate_ett}
\end{figure}
%\newline
% \\
% \textbf{Realistic ET Tubes Generation:}
\subsubsection{Realistic ET Tubes Generation:}
%\label{sssec:drawett}
We present next our methodology for generating synthetic ET tubes for adult X-ray images. In our solution, we were 
inspired by the work of Yi et al. \cite{yi2018automatic} that generated synthetic catheters on pediatric X-ray images. 
%, we used a similar method for generating synthetic ET tubes over adult X-ray images. 
Figure \ref{fig:generate_ett} depicts the ET tube generation steps:
First, we created a 2D simulation of the ET tube, as a hollow tabular object with a rectangular marker made of a radiopaque material. The tube and the marker are made from different materials and therefore have different attenuation components (c1 and c2).
The tube outer and inner width, d1 and d2, were chosen to fit an adult ET tube with strip thickness t. 
We defined $\{c1,c2,d1,d2,t\} = \{0.1,1,160,100,20\}$.
All the parameters above were selected 
%for generating the ET tube are derived 
based on  true physical properties of ET tubes or based on \cite{yi2018automatic}. 

In order to simulate the ET tube from different rotations, we projected the 2D profile using a Radon transform and sampled the projection at 0$^{\circ}$, 30$^{\circ}$, 60$^{\circ}$, 90$^{\circ}$. For each synthetic ET tube we selected one of the four profiles and sampled the values of 15 pixels for drawing the tube.
% $w=15$, the number of pixels for drawing the tube.

The trace of the ET tube was simulated over the trachea area using the clavicles segmentation. we extracted the middle point x between the clavicles and the lowest point y. Then, we randomly selected 4 points with x offset of $[-2,2]$ pixels and y-axis samples starting from 0 to y+offset of $[0,30]$ pixels.
The random points compose a line using B-spline interpolation. Finally, we draw the tube sampled profile over the line.
%in an anti-aliasing method \cite{wu1991efficient}.

The last step for creating a realistic X-ray with an ET tube is to merge the synthetic tube with the real X-ray image. We selected AP X-ray images from the NIH dataset that do not contain ET tubes and blended the random synthetic ET tubes into the images. We used a simple blending with random weights in the range of $[0.1,0.2]$.

\subsection{Detection and Segmentation CNN}
\label{ssec:detandsegcnn}

\begin{figure}[!t]
\centering
\includegraphics[width=4.6in]{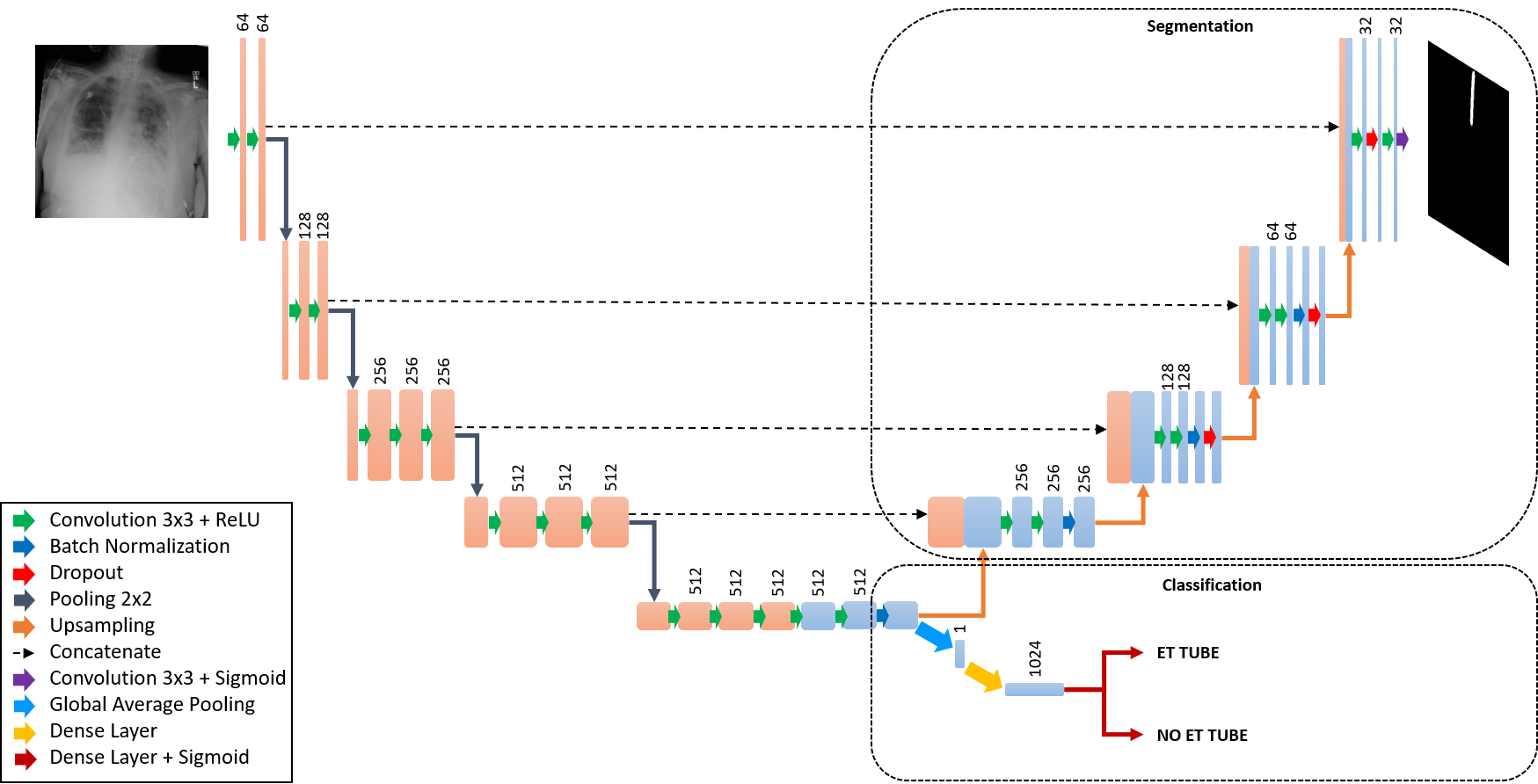}
\caption{ETT-Net: The proposed architecture for detection and segmentation of ET tubes}
\label{fig:arch}
\end{figure}

We propose a combined CNN architecture for ET tube detection and segmentation in chest radiographs, ETT-Net, as depicted in Figure \ref{fig:arch}. 
The architecture is built from a VGG16 style encoder followed by two paths: One is a decoder that continues the U-Net shape for addressing the ET tube segmentation task; The other path summarizes the features extracted at the end of the encoder using a global pooling layer followed by two dense layers and a sigmoid, for addressing the ET tube classification task. We used pre-trained VGG16 weights as initialization for the encoder.
The two paths of the network are trained simultaneously for both the classification task and the segmentation task using a combined weighted  Binary Cross-Entropy (BCE) loss and a Dice (D) loss as follows:
\begin{equation}
    \mathcal{L} = BCE(\hat{L},L) + \lambda D(\hat{S},S)
\end{equation}
where $L$ and $\hat{L}$ are the classification output label and the ground truth label, respectively;  S and $\hat{S}$ are the segmentation output mask and the ground truth mask, respectively.
$\lambda$ is the weight to balance between the loss components and was chosen (empirically) as 0.1.

The network inputs X-ray image of size $224\times224$ pixels duplicated 3 times (to fit the pre-trained encoder), the corresponding ET tube segmentation mask and a binary label for the presence or absence of ET tube. 
The input images are preprocessed with contrast limited adaptive histogram equalization (CLAHE) and normalized by their mean and standard deviation.
The segmentation masks can be a blank "all zero" image where no ET tube is present or a binary segmentation mask of the ET tube. For the training we augmented the data using horizontal flipping and small rotations of $\pm10^\circ$. 

\section{Experiments and Results}
\label{sec:expandres}

\subsection{Two Phase Training}
\label{sec:twostagetraining}

In order to train our suggested CNN using the synthesized data and still benefit from the existence of hundreds of X-ray images containing ET tube in the public dataset, we used a two phase training methodology. First, we trained the CNN using the generated data as explained in Section \ref{ssec:generatesynthdata}. Then, we used all the AP cases from the NIH dataset for inference: We extracted real cases to fine-tune the network to improve the classification and segmentation performance on real chest radiographs data.
In both training phases, we trained the network for 50 epochs using an Adam optimizer with default parameters.

The data for the first training phase includes 1669 X-ray images: 869 synthetic examples with ET tube and 800 without. The segmentation masks of the positive cases were obtained using a simple binary threshold of the synthetic tube before the blending operation. For the second phase, we used all NIH dataset AP cases and set conditions on the classification and segmentation outputs of the model: Images with classification prob. higher than 0.8 and non zero segmentation map were selected as positive examples; Images with classification prob. lower than 0.01 and a zero segmentation map were selected as negative examples. These conditions resulted in 3972 positive X-ray images with ET tube, and 36557 X-ray images without ET tube. Overall, after balancing the data, we trained the second phase using 7944 real chest radiographs. 

\begin{figure}[!t]
    \centering
    \begin{minipage}{0.45\textwidth}
        \centering
        \includegraphics[width=1.0\textwidth]{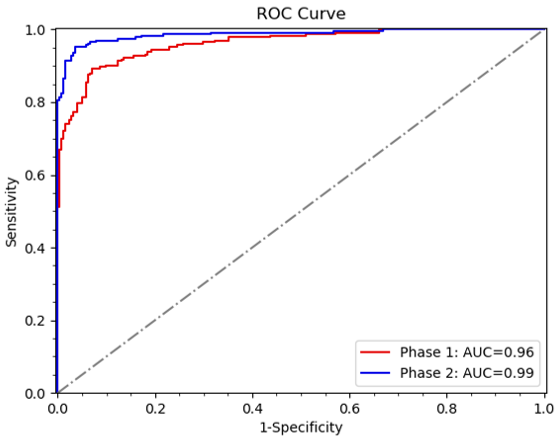} % first figure itself
        \caption{ROC curve after the two phases of training}
        \label{fig:roc}
    \end{minipage}\hfill
    \begin{minipage}{0.45\textwidth}
        \centering
        \includegraphics[width=0.9\textwidth]{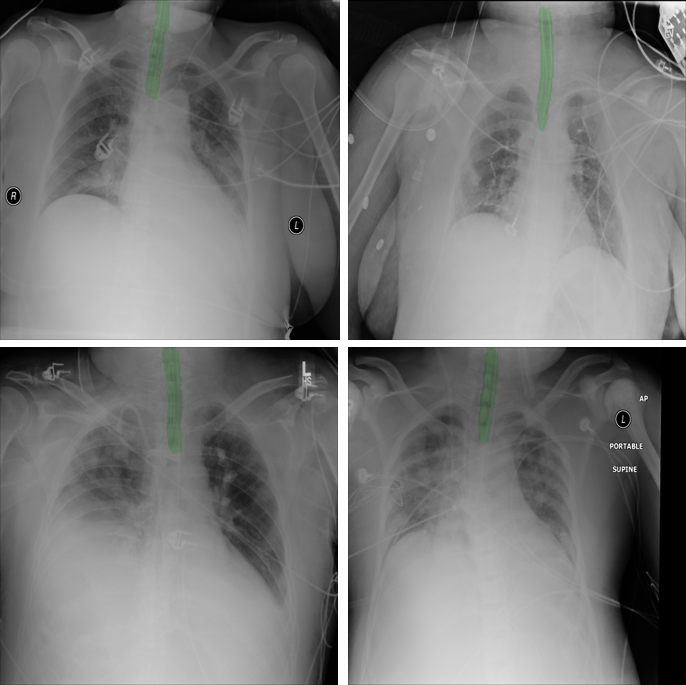} % second figure itself
        \caption{Cases correctly classified with ET tube and their corresponding segmentation as overlay (in green color)}
        \label{fig:results}
    \end{minipage}
\end{figure}

\begin{figure}[!b]
\centering
\includegraphics[width=4.7in]{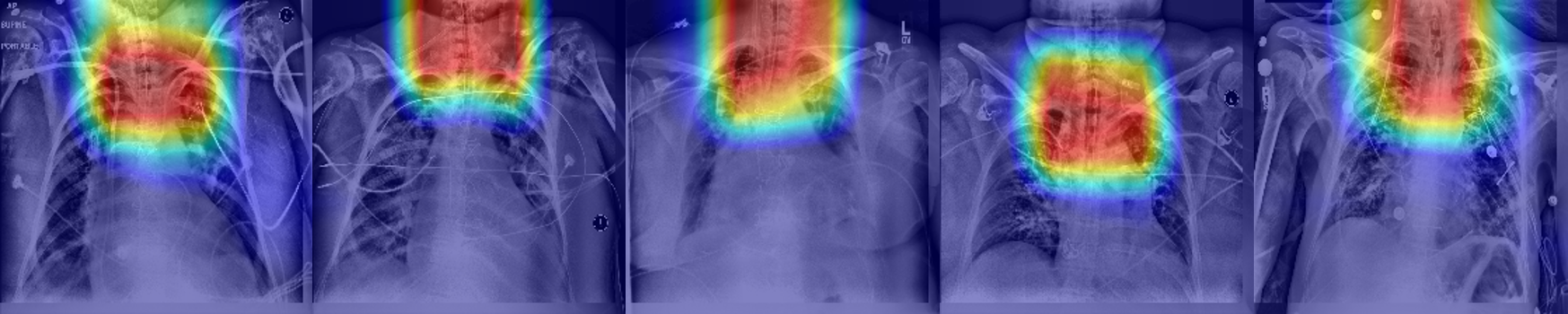}
\caption{Heatmap examples using ET tube classification CNN}
\label{fig:heatmap}
\end{figure}

\subsection{Test Set}
\label{sec:testset}

The test set includes 479 real chest radiographs from the NIH dataset that were collected manually one time during the development and entirely independent from all training data. All cases are in AP view position, 232 cases with ET tube and 247 without ET tube. After collecting the cases, we verified that the label for each case is consistent with the presence of an ET tube. It is important to note that as we didn't use manual annotations for the segmentation of the  tube, the ground truth segmentation maps are not pixel-wise accurate; still, they represent an expected range for  the ET tube position in the images. 
The classification accuracy is an quantitative measure we can use.  Thus, we tested our model using the AUC for the classification accuracy. The segmentation output was examined qualitatively.

% \begin{figure}[!t]
% \centering
% \includegraphics[width=2.7in]{ROC_curve.png}
% \caption{ROC curve after the two phases of training}
% \label{fig:roc}
% \end{figure}

% \begin{figure}[!t]
% \centering
% \includegraphics[width=4.7in]{result_examples5.png}
% \caption{Cases correctly classified with ET tube and their corresponding segmentation as overlay (in green color)}
% \label{fig:results}
% \end{figure}

\begin{table}[!t]
\caption{Comparison to state-of-the-art methods for classification between presence vs. absence of ET tube ; "-" means that the score was not reported}
\label{tabel:state_of_the_art_compare}
\centering
\begin{tabular}{p{4cm} p{1.1cm} p{3.5cm} p{3.2cm}}
\hline
{} & AUC & Sensitivity, Specificity & Testing Size [pos, neg] \\
\hline
% Ramakrishna et al. \cite{ramakrishna2011catheter} & - & 73.7\%, 91.3\% & 25 [19, 6] \\
Ramakrishna et al. \cite{ramakrishna2012improved} & - & 92.9\%, 97.2\% & 64 [28, 36] \\
Chen et al. \cite{chen2016endotracheal} & 0.95 & - & 87 [44, 43] \\
Lakhani et al. \cite{lakhani2017deep} & 0.99 & - & 60 [30, 30] \\
DenseNet  & 0.97 & 89.2\%, 93.0\% & 479 [232, 247] \\
ETT-Net - Phase1 & 0.96 & 89.2\%, 93.0\% & 479 [232, 247] \\
ETT-Net - Phase2 & 0.99 & 95.5\%, 96.5\% & 479 [232, 247] \\
% \\[-0.7em]
\hline
\end{tabular}
\end{table}

\subsection{Results}
\label{sec:results}

Training the combined model for classification and segmentation of ET tubes on synthetic X-ray images, we reached an AUC of 0.962 in classification accuracy. Using fine-tuning on real X-ray images, the accuracy improved to an AUC of 0.987 with both sensitivity and specificity over 95\% (Figure \ref{fig:roc}).
Figure \ref{fig:results} shows real chest radiographs from the test set that were classified correctly for presence of ET tube and their output segmentation maps.

We conducted an additional experiment using a different CNN architecture only for the classification task: identification  of the  tubes in real case scenarios. We trained a DenseNet \cite{huang2017densely} architecture with the same dataset we used in the second phase of the combined model - real cases with and without ET tube (n=7944) for 50 epochs and Adam optimizer. Training only for classification using a large real training data, we reached a high accuracy with an AUC of 0.975. Figure \ref{fig:heatmap} shows a heatmap visualization of the last convolutional layer of the network. This visualization clearly indicates the localization on   the ET tube area.  

Table \ref{tabel:state_of_the_art_compare} compares state-of-the-art methods for the classification of presence or absent ET tube to our methods - results after the first phase of training using the ETT-Net model, Second phase results of the final model after fine-tuning with real examples and the only classification method using DenseNet.
The table shows the amount of testing images used for testing each method, with separation for cases with ET tube (pos) and without (neg). Our best model, ETT-Net - Phase2, reached high performance with a test set size of one magnitude more than state-of-the-art methods. In addition our model was trained and tested using free public dataset (of real ICU patients) without the need for manual annotations, in contrast with the other methods were the cases were hand picked and annotated.

\section{Conclusion}
\label{sec:conclusion}

In this work, we proposed an approach for training a combined deep learning network for the tasks of detection and segmentation of the ET tube in adult chest radiographs without collecting and annotating data.
We used a public dataset of X-ray images and synthesized realistic ET tubes blended into those images. We used the synthetic data as a first phase of training our model. Collecting real X-ray cases using the trained model, we continued to a second phase of training. 
Both stages are trained using the ETT-Net - a combined CNN architecture for ET tube detection and segmentation in chest radiographs.
The combined model achieved a very high accuracy for the presence of ET tube in real ICU patients (0.99 AUC) using a test set which is ten times larger compared to previous studies and also outputs high quality segmentation maps that can assist in detection of the misplacement of the tubes. 
We also showed accurate results (0.97 AUC) using a CNN for classification only where the synthetic cases are used only for retrieval of real cases from the public dataset.
Future work can include exploring a similar method for other tube types and combining them together in a multi-class detection and segmentation method. 
The ideas presented in our paper for synthesizing data over public dataset images, can be used in other medical imaging domains (for example generating tumors over healthy patients in X-ray or CT studies).

% We demonstrated the results on a large test set (ten times larger compared to previous studies). We also showed accurate results (0.97 AUC) using a CNN for classification only.
% The ideas presented in our paper for synthesizing data over public dataset images, can be used in other medical imaging domains. 
%We hope that our ET tube detection and segmentation methods will be %incorporated into CAD systems and will help radiologists to save %precious time in the ICU for checking the ET tubes. 

% \bibliography{refs}

\end{document}